# Elucidating the Voltage Controlled Magnetic Anisotropy

Jia Zhang,[1,*] Pavel V. Lukashev,[2] Sitaram S. Jaswal,[1] and Evgeny Y. Tsymbal[1,**]

[1] Department of Physics and Astronomy & Nebraska Center for Materials and Nanoscience, University of Nebraska, Lincoln, Nebraska 68588, USA

[2] Department of Physics, University of Northern Iowa, Cedar Falls, Iowa 50614, USA

Voltage controlled magnetic anisotropy (VCMA) is an efficient way to manipulate the magnetization states in nanomagnets, promising for low-power spintronic applications. The underlying physical mechanism for VCMA is known to involve a change in the $d$-orbital occupation on the transition metal interface atoms with an applied electric field. However, a simple qualitative picture of how this occupation controls the magnetocrystalline anisotropy (MCA) and even why in certain cases the MCA has opposite sign still remains elusive. In this paper, we exploit a simple model of orbital populations to elucidate a number of features typical for the interface MCA and the effect of electric field on it, for 3d transition metal thin films used in magnetic tunnel junctions. We find that in all considered cases including the Fe (001) surface, clean $Fe_{1-x}Co_x(001)$/MgO interface and oxidized Fe(001)/MgO interface, the effects of alloying and electric field enhance the MCA energy with electron depletion which is largely explained by the occupancy of the minority-spin $d_{xz,yz}$ orbitals. On the other hand, the hole doped Fe(001) exhibits an inverse VCMA, where the MCA enhancement is achieved when electrons are accumulated at the Fe (001)/MgO interface with applied electric field. In this regime we predict a significantly enhanced VCMA which exceeds 1pJ/Vm. Realizing this regime experimentally may be favorable for a practical purpose of voltage driven magnetization reversal.

Magnetocrystalline anisotropy (MCA) is one of the critical parameters that determine the magnetization orientation, switching dynamics, and thermal stability of magnetic media. Ferromagnetic thin films with high magnetic anisotropy promise an easier operation scheme and higher signal stability. However, the large coercivity of these materials require high electric currents either to generate a magnetic field or produce a spin-transfer torque to write the bit information, which imposes significant limitations on their application in portable and high speed electronic devices. For example, in MgO-based magnetic tunnel junctions (MTJs), which are nowadays used in magnetic random access memories (MRAMs), current densities as large as $10^6$ A/cm$^2$ are required to reverse the magnetization of the magnetic bit by spin-transfer torque.

A possible approach to overcome this limitation is to use an electric field, rather than electric current, to control the magnetization of a ferromagnet.[1] Along these lines, voltage controlled magnetic anisotropy (VCMA) is considered as one of the most promising approaches to dramatically reduce the writing energy of MRAMs.[2] A large VCMA effect (~ 1 pJ/Vm) is, however, required for device application to overcome a large coercively of the ferromagnetic film which is needed for its thermal stability.

The effect of an electric field on the surface (interface) magnetic anisotropy has been extensively studied both theoretically[3-13] and experimentally.[14-24] Very large changes in magnetic anisotropy (VCMA ~ 5 pJ/Vm) were observed due to ionic motion and chemical reactions driven by applied voltage.[25-26] Ionic motion however is a slow process and can hardly be used for device applications where an ultrafast switching of magnetization is required. On the other hand, if the VCMA is caused purely by electronic effects, the timescale of the VCMA would lie on sub ns regime desirable for applications. It is known that VCMA in Fe(Co)/MgO-based MTJs is largely controlled by the effect of electric field screening on the ferromagnetic metal surface (interface) resulting in electron doping induced by an external electric field.[3]

It is understood that the main driving mechanism for VCMA is a change in the 3$d$-orbital occupation on the transition metal interface atoms with an applied electric field. Theoretical calculations are capable to predict reasonably well electric field induced changes in the surface MCA energy (MAE). However, an intuitive picture for VCMA is missing, although such a qualitative picture would be very helpful for the experimentalists to design materials and interface structures with enhanced VCMA.

In this paper, we consider a simple model which allows us to qualitatively explain the surface (interface) magnetic anisotropy and the effect of electric field on it. For 3$d$ transition metal thin films used in MTJs. We show that results of our first-principles density functional theory (DFT) calculations for the Fe (001) surface, clean $Fe_{1-x}Co_x(001)$/MgO interface and oxidized Fe(001)/MgO interface, involving effects of alloying and electric field, can be understood in terms of changes in the population of the minority-spin $d_{xz,yz}$ orbitals which enhance (reduce) the surface MAE with electron depletion (accumulation) at the interface. On the other hand, the hole doped Fe(001) exhibits an inverse VCMA, where the MCA enhancement is achieved when electrons are accumulated at the Fe(001)/MgO interface. In this regime, we predict a significantly enhanced VCMA which exceeds 1pJ/Vm. These results may be important for the search of the materials structures with enhanced VCMA which is critical for device applications.

The microscopic origin of MCA is the relativistic spin-orbit coupling (SOC), $\xi \mathbf{L} \cdot \mathbf{S}$, where $\mathbf{L}$ and $\mathbf{S}$ are the orbital and spin momentum operators respectively, and $\xi$ is the SOC constant. For not too thick films, the largest contribution to MCA comes from surfaces or interfaces due to the reduced symmetry. The surface (interface) MAE is known to be very sensitive to the local environment and details of the electronic band structure.

For 3$d$ transition metal thin films, the MAE can be evaluated using the second-order perturbation theory,[27-28]

which is applicable due to the relatively small SOC constant $\xi$ (typically in the range of 10-100 meV per atom) compared to the band energies, the crystal field splitting, and the exchange coupling. We define the MAE as the energy difference between the magnetization pointing along the $x$ direction in the plane of the film and the $z$ direction perpendicular to the plane, so that the positive MAE implies the out-of-plane easy axis, known as perpendicular magnetic anisotropy (PMA). Within the second-order perturbation theory the MAE is determined by the matrix elements of SOC between occupied and unoccupied states so that

$$E_{MAE} = \xi^2 \sum_{\substack{o,u \\ \sigma,\sigma'}} \frac{\left|\langle \psi_o^\sigma | L_z | \psi_u^{\sigma'} \rangle\right|^2 - \left|\langle \psi_o^\sigma | L_x | \psi_u^{\sigma'} \rangle\right|^2}{\varepsilon_u^{\sigma'} - \varepsilon_o^\sigma}, \quad (1)$$

where $\psi_o^\sigma$ and $\psi_u^{\sigma'}$ are unperturbed wave functions for occupied and unoccupied states with energies $\varepsilon_o^\sigma$ and $\varepsilon_u^{\sigma'}$ and spin states $\sigma$ and $\sigma'$, respectively.

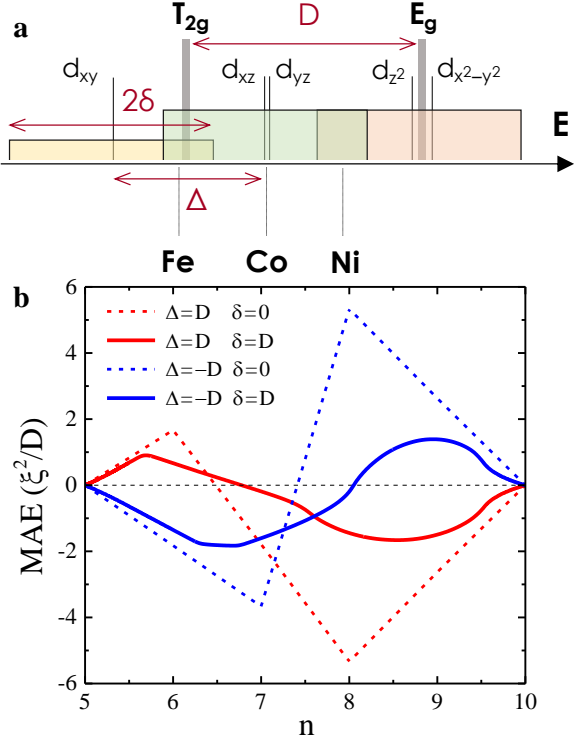

**FIG. 1.** (a) Schematic of the minority $3d$ orbital splitting by the crystal field of tetragonal symmetry. Broadening of the orbital levels mimics the energy bands. The vertical black dotted lines represent the Fermi energies of Fe, Co and Ni. (b) Magnetocrystalline anisotropy energy (MAE) as a function of orbital band filling $n$, for different values of the $T_{2g}$ level splitting $\Delta$ and broadening $\delta$.

For transition metal ferromagnets, such as bcc Fe, the MAE is determined by the energy bands formed from the $3d$ orbitals and depends on their occupations. The crystal field of cubic symmetry splits five $d$ orbitals into the $E_g$ doublet and the $T_{2g}$ triplet, as shown in Fig. 1a. At the (001) interface, the crystal field symmetry is reduced to tetragonal, which further splits these levels: $E_g$ into two singlets $A_1$ ($d_{z^2}$) and $B_1$ ($d_{x^2-y^2}$), and $T_{2g}$ into singlet $B_2$ ($d_{xy}$) and doublet $E$ ($d_{xz}, d_{yz}$). Fig. 1a shows schematically the orbital order as a function of the energy typical for the Fe/MgO (001) interface. The exchange splitting in Fe is sufficiently large, so that the majority-spin band is nearly fully occupied (see Fig. 2S-a in the Supplementary Material). We therefore ignore for simplicity contribution to the MAE from the majority-spin band. The dashed red line ($\Delta = D$, $\delta = 0$) in Fig. 1b shows the MAE as a function of orbital filling $n$ in Fig. 1a. It is seen that first the MAE is growing due to population of the $d_{xy}$ orbitals resulting in positive contribution to MAE in Eq. (1) through the $\langle d_{xy} | L_z | d_{x^2-y^2} \rangle$ matrix element. Then, it drops down and becomes negative, due to the negative contribution from the $d_{yz}$ orbitals[29] through the matrix elements $\langle d_{yz} | L_x | d_{z^2} \rangle$ and $\langle d_{yz} | L_x | d_{x^2-y^2} \rangle$. Finally, the MAE grows again due to filling of the $E_g$ states, reducing the latter contribution. Broadening of the orbital states that mimics the energy bands, as indicated in Fig. 1a, diminishes the sharp features of MAE versus $n$ but preserves the qualitative behavior, as seen from the solid red line in Fig. 1b.

There are a number of important implications which follow from this simplistic consideration. First, it is seen that the largest positive MAE (PMA) occurs for $n$ close to 6, corresponding to Fe. It is worth mentioning that the large PMA does not necessarily require hybridization of Fe-$d_{z^2}$ and O-$p_z$ orbitals across the Fe/MgO interface, as is often thought,[30-31] and can be explained purely based on the orbital population. In particular, our first-principles calculations predict a sufficiently large PMA of about 0.87 mJ/m$^2$ for the clean Fe (001) surface with no MgO. Second, the MAE drops when moving from Fe to Co in terms of the band filling and it becomes negative for Co. This is consistent with the results of DFT calculations performed for Fe$_{1-x}$Co$_x$/MgO interface as a function of $x$.[32] Third, since the VCMA is controlled by a change in the $3d$ band population produced by applied electric field, the slope of the curves in Fig. 1b determines the sign of VCMA. It is seen that for $n$ changing from 6 to 7 (i.e., from Fe to Co) the slope is negative, indicating that the MAE decreases with adding electrons, which is due to the population of the $d_{yz}$ orbitals. This is consistent with our DFT calculation for the Fe/MgO (001) interface, showing that the MAE increases with the increasing electric field pointing from Fe to MgO (electron depletion), as evident from Fig. 2a.[33,34] The increase of PMA with electron depletion is typical for most of the experiments on Fe(Co)/MgO interfaces[15-24] and is also known for the clean Fe (001) surface.[3] The dominant contribution comes from the $d_{yz}$ band, which is again consistent with the qualitative picture presented above. Finally, it is notable from Fig. 1b that reducing $n$ below 6 leads to the decrease of MAE, due to the reduction of the positive contribution from the $d_{xy}$ orbitals. This changes the sign of VCMA and has important implications as discussed below.

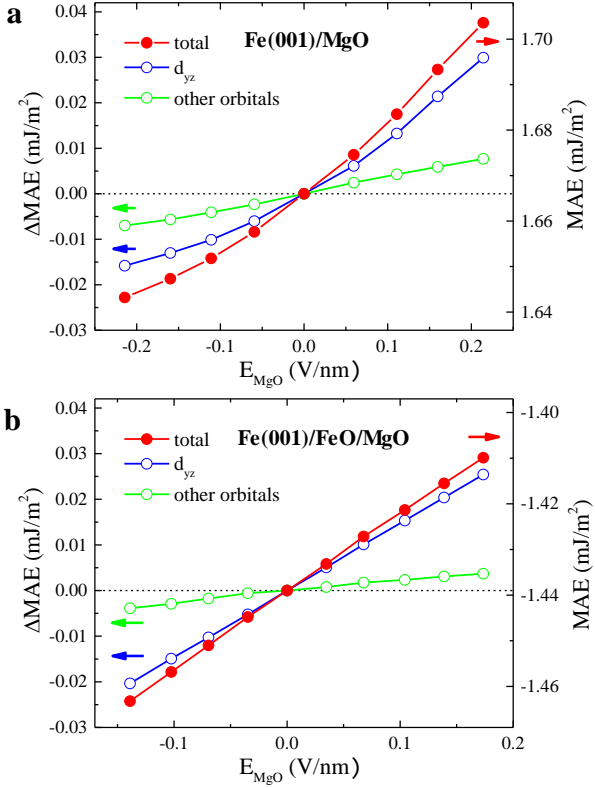

**FIG. 2.** Results of DFT calculations for Fe(001)/MgO (a) and Fe(001)/FeO/MgO (b) interfaces. MAE as a function of applied electric field $E_{MgO}$ in MgO (red curves) and $E$-field induced changes in the orbital contribution to MAE (green and blue curves). In the plot the electric field $E_{MgO}$ is scaled according to the experimental dielectric constant of MgO $\varepsilon = 9.5$.

Furthermore, this simple analysis explains a change in sign of the MAE, when Fe/MgO interface becomes oxidized. Our calculations predict that the MAE alters from about +1.67 mJ/m$^2$ for the clean Fe/MgO interface to –1.44 mJ/m$^2$ for the Fe(001)/FeO/MgO interface, where the first monolayer of Fe is fully oxidized (Fig. 2b).[33] This is due to the hybridization of the $d_{xy}$ orbitals of Fe and $p_{x,y}$ orbitals of O in the (001) plane, resulting in the formation of the bonding and antibonding states. The minority-spin antibonding state is largely composed of the $d_{xy}$ orbitals and pushed up in energy, as seen from Fig. 2S-b. This behavior is captured by our simple model with negative $\Delta$, which puts the $d_{xz,yz}$ doublet at the lowest energy. In this case, for $n$ changing from 5 to 7, the positive contribution to MAE from the $d_{xy}$ orbitals is eliminated, resulting in the increasing negative contribution coming from population of the $d_{xz,yz}$ states (Fig. 1b, blue curves). Moreover, this qualitative analysis predicts the VCMA sign consistent with our DFT calculation, showing that the MAE increases with the increasing electric field pointing from Fe to MgO and is largely determined by the $d_{yz}$ contribution (Fig. 2b).

Now we expand these considerations to elucidate the effects of electrostatic and chemical doping on VCMA of the Fe(001)/MgO interface. First, we consider the effect of electrostatic doping, assuming an ultrathin Fe layer which is typical for PMA experiments. A particular calculation is performed in a symmetric geometry of the MgO(3MLs)/ Fe(3MLs)/MgO(3MLs)/vacuum(2nm) supercell. In the calculation we vary the number of valence electrons by adding a valence charge to the supercell and neutralizing it by background of the constant charge density of opposite sign. Through the self-consistent electronic structure calculation, the valence charge in vacuum and in MgO is reassembled in a way to minimize the electrostatic energy, i.e. it is deposited on the Fe/MgO interface. The associated electric field emerges in vacuum and in MgO. This is equivalent to the calculation in an external electric field, apart from the constant background charge. The effect of the latter on the Fe metal is minimized due to the sufficiently large size of the supercell (3.91 nm) compared to Fe layer thickness (0.43 nm). Due to the imposed symmetry, the two interfaces in the supercell are exposed to the same field.[33]

The results of the calculation are shown in Fig. 3a, where MAE is plotted against the excess valence charge $n_e$. For $n_e = 0$, we find that MAE is about 1.57 mJ/m$^2$. It increases with adding holes to the system corresponding to an increasing electric field pointing from Fe to MgO. This behavior is consistent with the previous calculations[6] and typical for experiments[15-18]. For small values of $n_e$, the MAE changes linearly with $n_e$. The slope of the curve determines VCMA which is shown in Fig. 3b. Assuming that the electric field in MgO is given by $E_{MgO} = -n_e/\varepsilon\varepsilon_0 a^2$, where $\varepsilon$ is the dielectric constant of MgO, $\varepsilon_0$ is the vacuum permittivity and $a$ is the in-plane lattice parameter, we find the VCMA of about 0.25 pJ/Vm. This value is comparable to that in Fig. 2a (~0.15 pJ/Vm). Interestingly, with increasing $n_e$ up to about 0.025 ($E_{MgO} \approx -0.5$ V/nm) the MAE curve becomes flatter (also seen in Fig. 2a) and at $n_e \approx 0.03$ the VCMA is reduced. Such a non-linear variation of MAE (inset in Fig. 3a) has been seen in a number of experiments[16,18]. This behavior can be attributed to the resonant minority-spin interface state which is well known from spin-polarized tunneling.[35] In our calculation the interface state appears at about 0.05 eV above the Fermi energy and is largely composed of the $d_{xz,yz}$ orbitals (see, e.g., Fig. 2S-a). According to Plucinski et al.,[36] this state lies in the energy gap of the projected minority-spin bulk bands. Thus, electric field induced occupation of this state does not lead to a sizable reduction in the MAE (as expected for the $d_{yz}$ orbital) due to the absence of the unoccupied counterpart $E_g$ states to affect the MAE through $\langle d_{yz}|L_x|d_{z^2}\rangle$ and $\langle d_{yz}|L_x|d_{x^2-y^2}\rangle$ matrix elements.

The most striking feature, however, which is evident from Figs. 3a,b is the maximum in MAE at $n_e \approx -0.05$ and a change of VCMA sign at $n_e < -0.05$. This behavior is consistent with our simple analysis, according to which the inverse VCMA occurs when $n < 6$, due to the reduced contribution from the $d_{xy}$ orbitals (Fig. 1b, red lines). It is notable that in this inverse regime, the VCMA reaches a very large negative value of about –1.1 pJ/Vm at $n_e = -0.075$. The electric field in MgO corresponding to the VCMA maximum is $E_{MgO} \approx 1.73$ V/nm. Realizing this regime experimentally may be favorable for a practical purpose of voltage driven magnetization reversal.

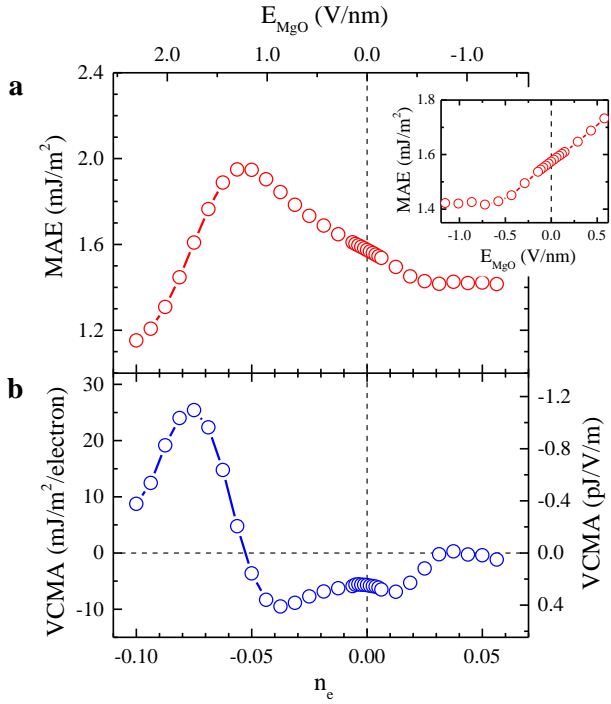

**FIG. 3**. Calculated MAE (a) and VCMA (b) as a function of the excess number of valence electrons $n_e$ (holes for negative $n_e$) at the Fe(001)/MgO interface. The inset shows the MAE as a function of electric field in MgO. The electric field is estimated from $E_{MgO} = n_e /\varepsilon\varepsilon_0 a^2$. The positive field is assumed to be pointing from Fe to MgO and corresponds to electron depletion.

The inverse VMCA may be achieved through the electrostatic doping caused by a top metal layer covering the PMA ferromagnet. Due to the different work functions of the PMA ferromagnet and the adjacent metal layer, charge transfer occurs between the layers to equalize the chemical potentials, changing the number of valence electrons in the ferromagnetic metal. The effect may be sizable only for ultrathin magnetic films due to the short screening length in metals. This may explain the results of Shiota et al.,[24] who observed a change of VCMA from normal to inverse, when Ta underlayer was replaced with Ru in the M/CoFeB/MgO (M = Ta, Ru) multilayer. The work function of Ta is about 0.25 eV lower than that of Fe, whereas the work function of Ru is 0.21 eV higher[37]. Therefore, it is expected that Fe is electron doped at the interface with Ta, but hole doped at the interface with Ru which may lead to the inverse VCMA. Also, the results of Nozaki et al.[18] demonstrate that VCMA is enhanced up to 0.29 pJ/Vm when a sub-nm thick layer of Fe is interfaced with Cr. At such a small thickness the effects of intermixing may play a role, resulting in the hole doping of Fe. According to Fig. 3b, a small hole doping raises VCMA up to about 0.4 pJ/Vm (at $n_e = -0.04$), which may be the origin of the enhanced VCMA observed by Nozaki et al.

Chemical doping is another way to manipulate VCMA. Here we consider effects of Fe alloyed with Co or Cr. Doping with Co adds electrons to Fe, whereas doping with Cr adds holes. This is expected to have an opposite effect on the MAE of the $Fe_{1-x}M_x$/MgO (M = Co, Cr) interface according to the simple analysis of Fig. 1b. DFT calculations are performed utilizing the coherent potential approximation (CPA) within the full relativistic screened-Korringa-Kohn-Rostoker (KKR) method.[33] The results are shown in Fig. 4, where the MAE is plotted against the number of doped electrons (holes), assuming that $n_e = x$ for Co, and $n_e = -2x$ for Cr. In agreement with the published results[32] the MAE is decreasing when adding electrons to Fe (FeCo/MgO). On the other hand, when Fe is doped with holes (FeCr/MgO) the MAE is decreasing. This is consistent with our expectation, though a simple picture of doping fails at a very small doping level where according to Fig. 3 we see an increase of MAE in the range of $n_e$ down to $-0.05$. This discrepancy indicates that doping with Cr is not exactly the same as the rigid decrease of number of electrons in Fe. Nevertheless, we see again a qualitative agreement between our simple picture presented in Fig. 1b and the accurate calculation shown in Fig. 4.

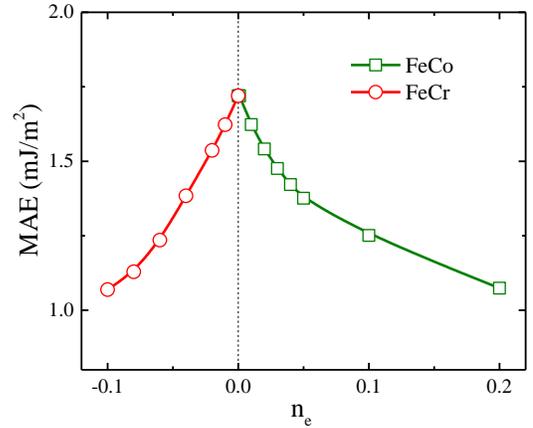

**FIG. 4**. MAE of the $Fe_{1-x}M_x$/MgO interface (M = Co, Cr) versus number of doped electrons $n_e$ (holes for negative $n_e$) calculated by KKR-CPA method.

In conclusion, starting from a very simple picture of orbital population, we have qualitatively explained the available results of the voltage controlled surface (interface) magnetic anisotropy of $3d$ transition metal thin films used in MTJs. The results of the DFT calculations for the Fe (001) surface, clean $Fe_{1-x}Co_x$(001)/MgO interface and oxidized Fe(001)/MgO interface, involving effects of alloying and electric field, can be understood in terms of changes in the population of the minority-spin $d_{xz,yz}$ orbitals which enhance the surface MAE with electron depletion at the interface. On the other hand, the hole doped Fe(001) exhibits an inverse VCMA, where the MCA enhancement is achieved when electrons are accumulated at the Fe(001)/MgO interface. In this regime, we predict a significantly enhanced VCMA which exceeds 1pJ/Vm. These results may be important to find the materials structures with enhanced VCMA which is critical for device applications.

This research was supported by the Nanoelectronics Research Corporation (NERC), a wholly owned subsidiary of the Semiconductor Research Corporation (SRC), through the Center for Nanoferroic Devices (CNFD). Computations were performed at the University of Nebraska Holland Computing Center.

* E-mail: kevinjiazhang@gmail.com
** E-mail: tsymbal@unl.edu

# SUPPLEMENTARY MATERIAL

# Elucidating the Voltage Controlled Magnetic Anisotropy

Jia Zhang,[1] Pavel V. Lukashev,[2] Sitaram S. Jaswal,[1] and Evgeny Y. Tsymbal[1]

[1] *Department of Physics and Astronomy & Nebraska Center for Materials and Nanoscience, University of Nebraska, Lincoln, Nebraska 68588, USA*

[2] *Department of Physics, University of Northern Iowa, Cedar Falls, Iowa 50614, USA*

## 1. DFT calculations for clean and oxidized Fe(001)/MgO interfaces

Density-functional theory (DFT) calculations are performed using projector augmented wave method (PAW),[1] implemented in the Vienna *ab initio* simulation package (VASP)[2] within the generalized gradient approximation (GGA) for the exchange-correlation potential.[3] The integration method[4] with a 0.05 eV width of smearing is used, along with a plane-wave cut-off energy of 500 eV and convergence criteria of $10^{-4}$ eV for ionic relaxations and $10^{-3}$ meV for the total energy calculations. The magnetocrystalline anisotropy energy (MAE) is evaluated as difference of energies calculated self-consistently corresponding to magnetization pointing along the [100] and [001] directions, in the presence of the spin-orbit coupling (SOC). The latter is included in VASP as a perturbation using the scalar-relativistic eigenfunctions of the valence states. The site-projected contributions to the MAE are calculated by evaluating the expectation values for the SOC energies: $E_{SOC} = \langle \frac{1}{c^2} \frac{1}{r} \frac{dV}{dr} \boldsymbol{L} \cdot \boldsymbol{S} \rangle$, where $c$ is the speed of light, $r$ is the distance in each atomic sphere, $V$ is the effective potential as a function of $\boldsymbol{r}$, and $\boldsymbol{L}$ and $\boldsymbol{S}$ are orbital and spin operators, respectively. For all considered systems, we use the experimental in-plane lattice constant of bcc Fe, $a = 0.287$ nm. The structures are relaxed in absence of electric field until the largest force becomes less than 5.0 meV/Å. Summation over k points is performed using a 24×24×1 mesh in the first Brillion zone, which according to our tests is sufficient to provide the calculated MAE accuracy of about 0.01meV. The site and orbital contributions to the MAE are obtained in VASP by evaluating the SOC energy difference between two magnetic axes ($\Delta E_{SOC}$).[5] Within this indirect approach it allows representing the MAE in terms of the sum over site and orbital contributions.

For the Fe(001)/MgO interface, in our calculation we use a MgO(4ML)/Fe(9ML)/MgO(4ML)/vacuum(2nm) supercell (Fig. 1S-a). The electric field is introduced using the dipole layer placed in the vacuum region of the supercell.[6] The positive electric field is defined as pointing away from the Fe layer to MgO. In this geometry we can evaluate effects of both positive and negative electric fields by performing calculations for only one direction of the field and evaluating the layer resolved contribution to the MAE at the two Fe/MgO interfaces. Typically, one or two interfacial monolayers of the ferromagnetic metal provide the dominant contribution to the MAE and VCMA due to the short-ranged electrostatic screening (Figs. 1S-b,c). In case of the oxidized Fe(001)/MgO interface we assume an additional oxygen atom which is placed in the first interfacial Fe monolayer atop the interfacial Mg atom, forming an Fe(001)/FeO/MgO interface.

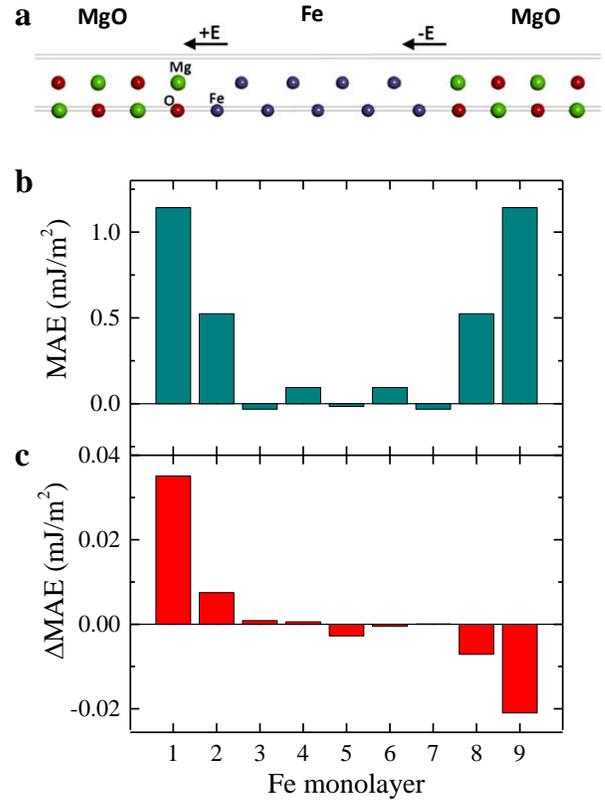

**Fig. 1S**: (a) MgO/Fe/MgO supercell structure (vacuum layer is not shown). (b,c) Contributions from different Fe sites (monolayers) to MAE (b) and to the change in MAE (ΔMAE) in applied electric field $E_{vac}$ = 2V/nm (c).

Figs. 2S-a,b show the orbital resolved density of states (DOS) at the interfacial Fe atom for the clean Fe(001)/MgO and oxidized Fe(001)/FeO/MgO interfaces, respectively. In the latter case, the interfacial Fe atom lies in the FeO plane. For the clean Fe(001)/MgO interface, there is a pronounced peak in the DOS of the interfacial Fe atom just above the Fermi energy. This peak is associated with the interface resonant state which is largely composed of the $d_{xz,yz}$ orbitals

(Fig. 2S-a, bottom panel, blue curve). It is notable that the sizable portion of the minority-spin $d_{xy}$ states are occupied for the clean interface (Fig. 2S-a, bottom panel, red curve), whereas for the oxidized interface these states are largely unoccupied and lie in the range of energies from 1 to 3 eV above the Fermi level (Fig. 2S-b, bottom panel, red curve). These are antibonding states resulting from the hybridization of the $d_{xy}$ orbitals of Fe and $p_{x,y}$ orbitals of O in the (001) plane.

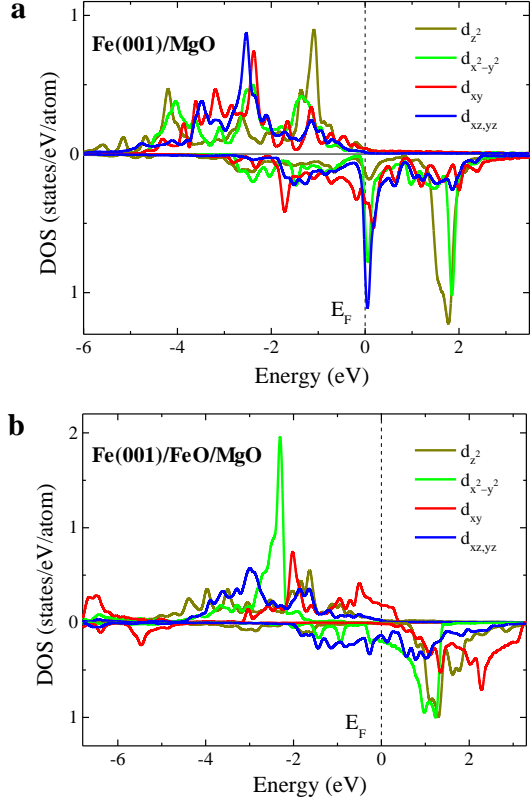

**FIG. 2S.** Orbital resolved density of states (DOS) at the interfacial Fe atom for clean Fe(001)/MgO (a) and oxidized Fe(001)/FeO/MgO (b) interfaces. Top and bottom panels correspond to majority- and minority-spin contributions, respectively.

Note that in all our calculations we do not relax the atomic structure in the in the presence of electric field. The electric field in MgO layer is expected to be $E_{MgO} = E_{vac}/\varepsilon$, where $\varepsilon$ is the dielectric constant of MgO and $E_{vac}$ is the electric field in vacuum. From our calculation we estimate the dielectric constant to be $\varepsilon \approx 3.3$ from the calculated ratio of potential slop between MgO and vacuum, which is less than the experimental value of $\varepsilon \approx 9.5$, due to the neglect of the ionic response of MgO in the calculation. In order to take into account this deficiency, we plot the MAE in Figs. 2a and 2c against the expected electric field in MgO corresponding to experimental conditions, i.e. $E_{MgO} = E_{vac}/9.5$.

## 2. DFT calculations for the "electrostatically doped" Fe(001)/MgO interface

Similar to the above, the first-principles calculations for the electrostatically doped Fe(001)/MgO interface are performed using VASP. A symmetric supercell MgO(3MLs)/Fe(3MLs)/MgO(3MLs)/vacuum(2nm) is used in the calculations. The structure is relaxed for the neutral system until the force on all atoms is less than 1 meV/Å. The MAE is evaluated using the force theorem.[7] Within this approach, first, the electronic structure is self-consistently calculated in the absence of SOC by using 16×16×1 k-points k-point grid in the Brillouin zone. Then, the MAE is obtained by taking the band energy difference in the presence of SOC between two magnetic axes along [100] and [001] directions with a finer 32×32×1 k-points mesh. Electrostatic doping is performed by changing the number of valence electrons in the whole system and neutralizing this charge by background of the constant charge of opposite sign. By using this method, the excess charge density in vacuum and MgO is redistributed in a way to deposit the most part of the charge to the metal surface (interface), which is analogous to the electric field effect. The constant charge density background extending to the metal region produces a "chemical" doping effect in addition to the "electrostatic" doping through the electric field. This effect is minimized in our calculations due to the sufficiently large size of the supercell (3.91 nm) compared to Fe layer thickness (0.43 nm). Atomic relaxations are not performed for the charged system. Fig. 3S shows results of calculation for the excess valence charge $n_e = -0.025$ (the negative number corresponds to adding holes). It is seen that the electrostatic potential in vacuum has a parabolic shape which is due to a constant background charge in vacuum. The MAE and $n_e$ for the Fe(001)/MgO interface (shown in Fig. 3 of the main text) is considered to be a half of the corresponding values for the supercell, which contains two identical interfaces.

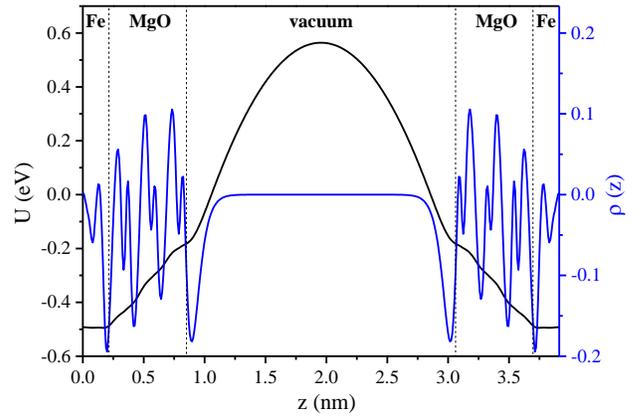

**Fig. 3S**: Calculated electrostatic potential energy (black curve) and excess charge distribution (blue curve) across the MgO(3MLs)/Fe(3MLs)/MgO(3MLs)/vacuum(2nm) supercell structure for $n_e = -0.025$.

### 3. DFT calculations for the chemically doped Fe(001)/MgO interface

Calculations of the MAE for the $Fe_{1-x}M_x$/MgO (M = Co, Cr) interfaces are performed using the full relativistic screened-Korringa-Kohn-Rostoker (KKR) method based on density functional theory, where the spin-orbit coupling is taken into account by solving the full relativistic Dirac equation.[8] The coherent potential approximation (CPA) is utilized to describe the compositional dependence of the $Fe_{1-x}M_x$ alloys. The potentials are described within the atomic sphere approximation (ASA). Particular calculations are performed using MgO(3MLs)/FeM(3MLs) supercell geometry by imposing periodic boundary conditions. More details of the calculations can be found in ref. 9.